\documentclass[3p]{elsarticle}
\usepackage{amssymb}
\usepackage{mathrsfs}
\usepackage{color}

\usepackage{hyperref}

\journal{Astroparticle Physics}
\begin{document}
\begin{frontmatter}

\title{On the prospects of cross-calibrating the Cherenkov Telescope Array with an airborne calibration platform.}

\author{Anthony M. Brown}
\address{Department of Physics and Centre for Advanced Instrumentation, University of Durham, South Road, Durham, DH1 3LE, UK}
\ead{anthony.brown@durham.ac.uk}

\begin{abstract}
Recent advances in unmanned aerial vehicle (UAV) technology have made UAVs an attractive possibility as an airborne calibration platform for astronomical facilities. This is especially true for arrays of telescopes spread over a large area such as the Cherenkov Telescope Array (CTA). In this paper, the feasibility of using UAVs to calibrate CTA is investigated. Assuming a UAV at 1km altitude above CTA, operating on astronomically clear nights with stratified, low atmospheric dust content, appropriate thermal protection for the calibration light source and an onboard photodiode to monitor its absolute light intensity, inter-calibration of CTA's telescopes of the same size class is found to be achievable with a $6-8$\% uncertainty. For cross-calibration of different telescope size classes, a systematic uncertainty of $8-10$\% is found to be achievable. Importantly, equipping the UAV with a multi-wavelength calibration light source affords us the ability to monitor the wavelength-dependent degradation of CTA telescopes' optical system, allowing us to not only maintain this $6-10$\% uncertainty after the first few years of telescope deployment, but also to accurately account for the effect of multi-wavelength degradation on the cross-calibration of CTA by other techniques, namely with images of air showers and local muons. A UAV-based system thus provides CTA with several independent and complementary methods of cross-calibrating the optical throughput of individual telescopes. Furthermore, housing environmental sensors on the UAV system allows us to not only minimise the systematic uncertainty associated with the atmospheric transmission of the calibration signal, it also allows us to map the dust content above CTA as well as monitor the temperature, humidity and pressure profiles of the first kilometre of atmosphere above CTA with each UAV flight.
\end{abstract}


\end{frontmatter}


\section{Introduction}

The Cherenkov Telescope Array (CTA) is the next generation of very high energy $\gamma$-ray instrument. Building on the strengths of the current ground-based $\gamma$-ray facilities, CTA is designed to achieve at least an order of magnitude improvement in sensitivity with unprecedented energy and angular resolution \cite{cta}. Importantly, compared to the current telescopes' energy range of $\sim50$ GeV to $\sim10$ TeV, CTA will also increase the energy reach of ground-based $\gamma$-ray astronomy, with an evisaged energy threshold of 20 GeV, extending to beyond 100 TeV. To realise these ambitious goals, CTA will comprise three telescope size classes: Large-Sized-Telescopes (LST), Medium-Sized-Telescopes (MST) and Small-Sized-Telescopes (SST), with each size class optimised to observe a different $\gamma$-ray energy range \cite{cortina,medium,sst}. With a diameter of $\sim23$~m, the LSTs are designed to observe the low Cherenkov photon intensity associated with $20 \leq E_{\gamma} \leq 200$~GeV photon-induced extensive air showers (EAS). The MSTs will have a diameter of $\sim 12$ m and provide the majority of the improvement in flux sensitivity in the $0.1 \leq E_{\gamma} \leq 10$~TeV energy range. The SSTs, in either single or dual mirror optics configuration, will extend the high-energy reach of CTA, predominantly observing $\gamma$-rays with energies from a few TeV up to 100 TeV and beyond. To allow all-sky coverage, the CTA observatory will consist of two arrays, one in each hemisphere. The northern array is intended to contain approximately 20 telescopes (LSTs and MSTs) spread over about 1 km$^2$, while the southern array is intended to contain $\sim100$ telescopes (LSTs, MSTs and SSTs) spread over an area of approximately four square kilometres. More than two thirds of the southern array telescopes are foreseen to be SSTs. Furthermore, the enrichment of the southern array is envisaged with the addition of medium-sized dual-mirror Schwarzschild-Couder Telescopes (SCTs; \cite{scts}), which have the advantage of a better angular resolution, smaller point spread function and a larger field of view compared to the traditional single-mirror Cherenkov telescopes. 

CTA will herald a new era for ground-based $\gamma$-ray astronomy, with the emphasis shifting from source discovery, to population studies and precision measurements. This change in emphasis requires a reduction in the systematic uncertainties of the observing technique, compared to the current ground-based $\gamma$-ray facilities. In particular, to meet the science goals of CTA, the systematic uncertainty of the energy of a photon candidate (at energies above 50 GeV) must be $<$15\%, with $<$10\% in the systematic shift of the estimated energy scale. The uncertainty of the collection area of the system above 40 GeV for dark sky conditions must be $<$12\%, with a goal uncertainty of $<$8\%, and the absolute intensity measurement of the Cherenkov light (post-calibration) at the position of each telescope must be known with a systematic uncertainty $<$8\%, with a goal of 5\% \cite{daniels, gaug3}. If these goals are met, CTA will probably resolve many spectral features,  such as spectral cut-offs or spectral hardenings, which hitherto have been observed as statistically compatible with simple power-laws. To realise these systematic uncertainty goals, robust calibration of telescope response, including optical throughput, is required across the whole array. This is helped if a common, well-understood calibration light source can be used.

For practical reasons, ground-based Cherenkov telescopes have no housing for their structures and therefore are exposed to the elements and thus subject to the effects of weathering. These weathering effects contribute to the degradation of the telescopes' optical systems, such as mirrors. Since the energy of the initial $\gamma$-rays is deduced from the amount of Cherenkov light observed, an unmonitored change in the optical throughput of the telescope system can introduce a large uncertainty in the absolute energy determination for the $\gamma$-rays. Furthermore, CTA's different telescope size classes will employ different camera technology, particularly with regards to the lightsensors, such as photomultiplier tubes (PMT) (e.g. \cite{nectar,lstcam,flash}) and Silicon photomultipliers (SiPM) (e.g. \cite{astricam,sct,sst-1mcam,mespie}). These different lightsensors will have different spectral responses. As such, an unmonitored wavelength-dependence of the degradation of the optical systems will introduce an additional uncertainty since it will have a different impact on the observed signal depending upon which camera technology has been used.


Considerable effort has been made to investigate a variety of calibration methods for understanding the optical throughput of CTA's telescopes \cite{me,toscano,mineo,markusclf,rode,mitchell,parsons}. While an accuracy significantly better than 10\% is possible for all the methods investigated, there are a number of hardware requirements and operational limitations. Furthermore, to meet the CTA requirements on systematic uncertainties, the wavelength-dependency of the telescopes' optical throughput should be determined and monitored. This requires the regular use of a multi-wavelength (MWL) capable calibration light source to monitor the optical throughput of each individual CTA telescope at different wavelengths \cite{macca}. Given the large number of telescopes expected within CTA, combined with the CTA requirement that a minimal amount of possible observing time is used for calibration procedures, the use of a cross-calibration technique that simultaneously illuminates as many telescopes as possible is of great importance and must be pursued where possible. To address this need, a MWL-capable vertically-firing central laser facility (CLF) has been investigated \cite{markusclf}. However, while a MWL-capable CLF would be able to achieve a relative cross-calibration of CTA at the 5\% level, the hardware requirements for using a CLF, particularly with regards to signal readout for the camera, appear to be prohibitive because of the need to read out much longer pulses than those of the Cherenkov light from EAS. Importantly, the hardware requirements and operational limitations of the methods studied so far, combined with the need for MWL information on the optical throughput, can be addressed with a UAV-based calibration system.


The advances in battery performance, flight control software and carbon fibre technology over recent years have made the use of small unmanned aerial vehicles (UAVs) as an airborne calibration platform for astronomical facilities a possibility. Early work by the Pierre Auger Observatory (PAO) took advantage of these advances to use UAVs to calibrate individual pixels, and investigate the point spread function, of their fluorescence telescopes \cite{bauml}. In particular, an omnidirectional UV light source was attached below an octocopter UAV, and positioned $\mathcal{O}(0.5-1.0)$~km in front of a fluorescence telescope. At this distance, individual pixels of the telescope were illuminated. Expanding upon this early work, using the same calibration payload, the efficiency of the fluorescence detectors of the PAO and Telescope Array cosmic ray detectors were cross-calibrated \cite{matthews}, with the Telescope Array recently flying their own calibration flight platform \cite{hayashi}. Building upon the experience of the PAO, a UAV-based calibration system has also been used to characterise the far-field beam map of three CROME microwave reflector antennas \cite{felix}, and more recently for the beam map of a single radio dish \cite{chang}. 

An airborne calibration platform has the potential to perform a number of calibration procedures for a sparsely spaced telescope array such as CTA. The primary role envisaged for CTA's airborne calibration system are the cross-calibration of different telescope types by placing a well-known calibration light source at altitude above the array and monitoring the MWL performance of CTA telescopes' optical throughput. Nonetheless, the flexibility and versatility of the UAV platform allows for a wide variety of calibration and monitoring procedures to be performed. For example, in addition to the primary payload for cross-calibrating the optical throughput, a secondary payload of environmental sensors such as nephelometers, pressure, temperature and humidity sensors, allows the atmosphere above CTA to be characterised in detail. In particular, when used in conjunction with a LIDAR system, a UAV-based nephelometer can study the boundary layer region where aerosol mixing is most highly concentrated and where the molecular profile is influenced by ground effects.


This paper outlines the feasibility of using a UAV to cross-calibrate the optical throughput of the individual telescopes of CTA. In particular, the primary factors governing the achievable precision of a UAV-based calibration system are identified and a prediction is made of the uncertainty a UAV-based system can achieve if these factors are managed. The paper's outline is as follows: in \textsection2 the feasibility of cross-calibrating both telescopes of the same type and different telescope types is investigated. In \textsection 3 the accuracy of a UAV system is discussed, with the necessary hardware requirements to achieve this accuracy being outlined in \textsection 4. Finally, in \textsection 5 the advantages of using a UAV-based calibration system are discussed, particularly with regards to monitoring any wavelength-dependent degradation of the telescopes' optical systems with a MWL-capable flasher system. 

\section{Cross-calibrating the optical throughput of CTA's telescopes}
\label{crosscalib}

\begin{figure*}
 \centering
\includegraphics[width=120mm]{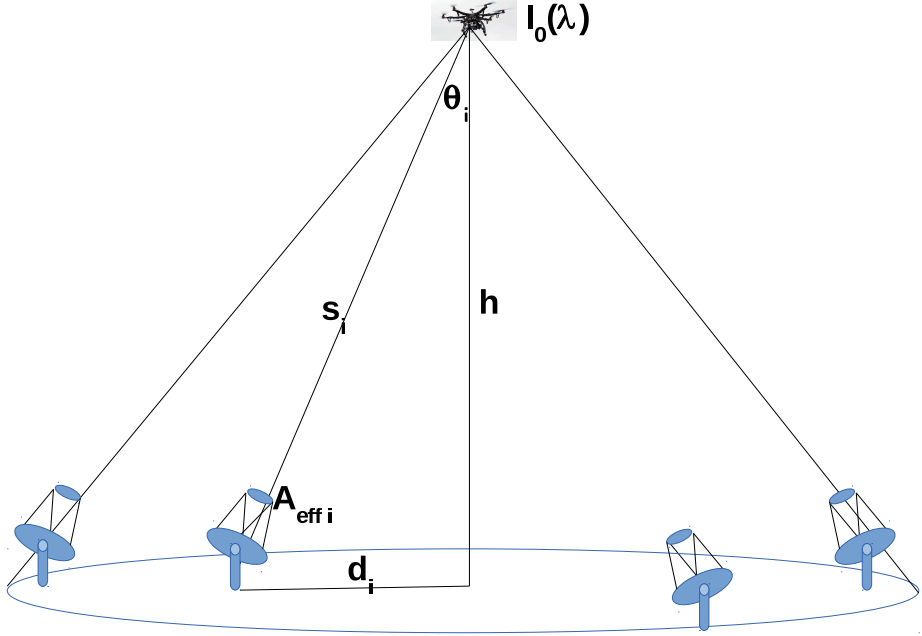}
\caption{A sketch of the baseline concept for using an airborne platform to cross-calibrate the telescopes of CTA. The UAV is flown at altitude $h$ above CTA, with the pathlength from the UAV to telescope $i$, given by $s_i$. $A_{\textit{eff}}$ is the effective area of the telescope and $I_{o}(\lambda)$ is the intensity of the calibration light source. $d_i$ is the horizontal distance from the telescope to the UAV.}
\label{baseline}
\end{figure*}

The airborne calibration system will consist of two main components: a UAV and a scientific payload. The UAV will consist of a flight platform, such as a hexa- or octocopter, which communicates with a ground station via a radio link. The scientific payload is envisaged to be modular in design to allow for a variety of calibration procedures, with minimal modification to the flight platform. Furthermore, this modular design allows for periodic calibration of the scientific payload in the lab, and affords us the ability to easily update or replace components of the scientific payload to minimise systematic uncertainties.

The baseline concept for using an airborne platform to cross-calibrate the telescopes of CTA is to position the UAV at a stable altitude above the array and illuminate as many telescopes as possible with a wide-beamed calibrated light source. The camera of each telescope captures the reflected light, ideally contained in one single pixel. The field of view (FoV) of CTA's pixels are telescope dependent. The smallest pixel FoV will be the SCT at $0.067^{\circ}$ \cite{sct}, with largest pixel FoV the single mirror MST with a pixel FoV of $0.18^{\circ}$ \cite{flash}. The LST will have a pixel FoV of $0.1^{\circ}$ \cite{cortina}. For the SSTs, the pixel FoVs will be $0.25^{\circ}$, $0.17^{\circ}$ and $0.15^{\circ}$ for the SST-1M, ASTRI and GCT telescopes respectively \cite{sst}. 

While the natural tendency would be to fly as high as possible to maximise the number of telescopes simultaneously illuminated by the UAV, there is a compromise that needs to be considered between the flight performance envelope of the UAV, such as flight time or payload weight, the performance of the light source payload and the maximum distance possible for a reliable radio telemetry link between the UAV and the ground station. Additionally, atmospheric extinction of the calibration light must be controlled to the few percent level. Nonetheless, throughout this feasibility study, the altitude of the UAV is considered a free parameter. The general geometry of this baseline concept is shown in Figure \ref{baseline}. 

Assuming a stable positioning of the UAV above CTA such that it is always in the FoV of a telescope, the amount of light received by an individual telescope is given by:

\begin{equation}
I(\lambda,s)= I_{o}(\lambda,\Omega) \cdot  \mathscr{T}_{aer}(\lambda,s) \cdot \mathscr{T}_{mol}(\lambda,s) \cdot (1+f) + F_{bgr},
\end{equation}

where $\lambda$ is the wavelength of the calibration light source used, $s$ is the distance from the UAV to the a given telescope, $\Omega$ is the solid angle of the calibration source's light pattern, $I(\lambda,s)$ is the intrinsic intensity of the light at the focal plane of the individual telescope, $I_{o}(\lambda,\Omega)$ is the intensity of the calibration light source in terms of the number of photons per steradian, $\mathscr{T}_{aer}(\lambda,s)$ and $\mathscr{T}_{mol}(\lambda,s)$ are the aerosol and molecular transmissions respectively, $(1+f)$ is a correction factor to account for multiple scatterings and $F_{bgr}$ denotes the ambient background light, from both the night sky and star light, detected during the signal integration window. If we assume that the calibration light source is constructed such that illumination is uniform throughout the considered solid angle, Eq. 1 can be simplified to:

\begin{equation}
I(\lambda,s)= I_{o}(\lambda) \cdot \frac{A_{\textit{eff}}}{s^2} \cdot \mathscr{T}_{aer}(\lambda,s) \cdot \mathscr{T}_{mol}(\lambda,s) \cdot (1+f) + F_{bgr},
\end{equation}

where $A_{\textit{eff}}$ is the effective mirror area of the telescope. While the effective mirror area for each telescope is initially calibrated, temporal variations of the properties such as mirror reflectivity results in the need to periodically quantify each telescope's effective area by monitoring their optical throughput.  As such, for cross-calibrating CTA, the telescope effective mirror area is the term to be calibrated. 

Unlike a central laser facility (e.g. \cite{augerclf,markusclf}), which must consider a summation of different atmospheric transmissions from the ground to the scattering altitude and back to the telescope, the total atmospheric transmission for a UAV-based light source, $\mathscr{T}_{atm}$, where $\mathscr{T}_{atm} = \mathscr{T}_{aer}(\lambda,s) \cdot \mathscr{T}_{mol}(\lambda,s)$, simply depends upon the distance of the UAV from the telescope for a given wavelength. The atmospheric transmission from the UAV located at altitude $h$ above the array, to an individual telescope is given by:

\begin{equation}
 \mathscr{T}_{atm}(\lambda,s) = \textrm{exp}[-\int^{h}_{0} (\kappa_{aer}(\lambda,s) + \kappa_{mol}(\lambda,s)) ~\mathrm{d}s],
\end{equation}

where $\kappa_{aer}(\lambda,s) $ and $ \kappa_{mol}(\lambda,s)$ are the aerosol and molecular extinction coefficients respectively. Assuming that the aerosol distribution is stratified such that it is horizontally uniform, for example see \cite{augerclf2}, Eq. 3 can be simplified to:  

\begin{equation}
 \mathscr{T}_{atm}(\lambda,\textrm{h},\theta) = \frac{\textrm{exp}(-\tau_{atm}(\lambda,h))}{\cos(\theta_i)},
\end{equation}

where $\tau_{atm}(\lambda,h) = \tau_{aer}(\lambda,h) + \tau_{mol}(\lambda,h)$ is the combined vertical optical depth of the UAV at altitude $h$ above CTA and $\theta_i$ is the angle from the UAV's nadir to telescope $i$ (see Figure 1). $\mathscr{T}_{atm}$ can be maximised by operating the UAV on clear nights where we expect extinction due to atmospheric dust content to be minimum. This criterion would also remove the multiple scattering component of Eq. 1, such that $(1+f) \simeq 1$. Combining Eq. 2 and Eq. 4, the amount of light received by a given telescope can be simplified to:

\begin{equation}
 I_{i} = I_{o}(\lambda) \cdot \frac{A_\textit{eff}}{s^2} \cdot \mathscr{T}_{atm}(\lambda,h,\theta) + F_{bgr},
\end{equation} 

It is important to note that, with the assumption of low aerosol content which is horizontally stratified, Eq. 5 is an approximation which is most accurate on clear nights with little aerosol content. The uncertainty introduced by the stratified aerosol content assumption is addressed in \textsection 3.

\subsection{Inter-calibration }
\label{secintra}

\begin{figure*}
 \centering
  \includegraphics[width=120mm]{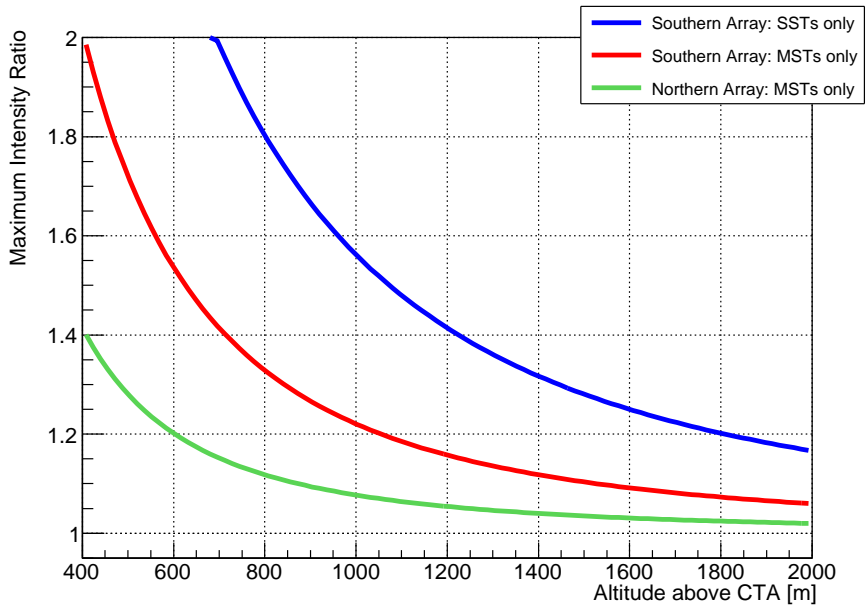}
  \caption{The ratio of the light intensity, as given by Eq. 5, observed by the telescope closest to the UAV compared to the furthest telescope, as a function of UAV altitude above CTA, when comparing the SST and MST telescopes. The array layout is taken from \cite{layout}. Note that the LST telescopes are envisaged to be positioned symmetrically around the centre of CTA arrays, as such, if the UAV is flown above the centre of the array, all LSTs are expected to receive the same amount of light.}
  \label{inten}
\end{figure*}

For calibrating telescopes of the same size class, hereby referred to as inter-calibrating, we ideally want the spread in the observed calibration signal amplitude across the illuminated telescopes to be as small as possible. This desire is primarily driven by the need to limit the possibility of camera saturation and hence disentangle saturation-induced non-linear effects of the camera's front-end electronics' response from the degradation of the telescopes' optical throughput. To investigate the maximum differences in signal amplitude across CTA from a UAV-mounted light source, the amount of light observed, as described by Eq. 5, for the telescope closest to the UAV is compared to that of the telescope furthest from the UAV for each telescope size class. Both telescopes are assumed to have the same flux of background light, $F_{bgr}$, in the recorded signal, conservatively estimated to be always smaller than 5 photoelectrons per pulse on average\footnote{In reality this background signal is telescope dependent, with the background light being an order of magnitude smaller for the SSTs (see \textsection2.2)}. Figure \ref{inten} shows the maximum differences when comparing telescopes of the same type, as a function of UAV altitude above CTA. 

Figure \ref{inten} shows a preference towards higher UAV operating altitude for minimising the difference in signal amplitude across the array. Taking a reference UAV altitude of 1300~m above CTA\footnote{Assuming a wide-beam calibration light source, utilising a diffuser with a $50^{\circ}$ opening angle, a UAV positioned at an altitude of 1300~m above CTA will be able to illuminate the entire CTA-North array in one positioning, and will take four positionings to illuminate the entire CTA-South array.}, for the MSTs at the northern site there is a $\sim5$\% difference in the signal amplitude when considering the closest and furthest telescopes from the UAV. For the MSTs on the larger southern site, there is a maximum difference in signal amplitude of $\sim14$\% when only one UAV positioning is considered directly above the centre of the array. A maximum signal difference of 14\% allows us to inter-calibrate the MSTs, at a high illumination level, without saturating the MST nearest to the UAV. For the SST component of CTA-South, where it is foreseen to be $\mathcal{O}(1200)$m between the central and border SSTs, we find a $\sim36$\% difference in the signal amplitude between the closest and furthest telescopes from the UAV. As such, even with the large physical separation, a difference of only 36\% in illumination level recorded across the SST component of CTA allows us to easily inter-calibrate the SSTs, at a high illumination level, without saturating the SST nearest to the UAV. The LST component of CTA is envisaged to be positioned symmetrically around the centre of the array. As such, if CTA is calibrated with one UAV positioning above the array centre, as Figure \ref{inten} suggests, there is no difference in the amount of light received by each of the LSTs.

\subsection{Cross-calibration}
\label{secinter}

For the calibration of telescopes of different size classes, hereby referred to as cross-calibration, the differences in telescope mirror area result in a larger range of signal amplitudes recorded across the array when compared to the simple geometrical arguments of the inter-calibration. Furthermore, the different mirror sizes result in a different amount of background light flux, $F_{bgr}$, for each telescope size class. To account for that, $F_{bgr}$ is calculated for each telescope size class separately, assuming each telescope is pointing to a dark region of the sky away from the Galactic and ecliptic planes. Typical background light measurements for the MAGIC telescope on La Palma have been found to be at the level of $0.13$ photoelectrons per pixel per nanosecond \cite{magicnsb}. Extrapolating this value, taking into consideration the expected field of view for CTA's LST, MST and SST telescopes, we expect $\sim0.4$, $\sim0.19$ and $\sim0.03$ photoelectrons per pixel per nanosecond for the LST, MST and SST structures respectively \cite{markusclf}. For a 5 ns calibration pulse these rates equate to 2.0, 1.0 and 0.15 background photoelectrons for the LST, MST and SST respectively \footnote{It should be noted that, higher background light levels are derived if one extrapolates from Preuss et al. 2002 \cite{preuss}.}. It should be noted that, while a 5 ns calibration pulse can be temporally elongated due to the telescope dish shape or the camera data aquisition system, and that some CTA telescopes may have a fixed signal integration window, the number of background photons measured by the cameras during operation can be determined routinely online with, for example, the use of random triggers.

\begin{figure*}
 \centering
  \includegraphics[width=120mm]{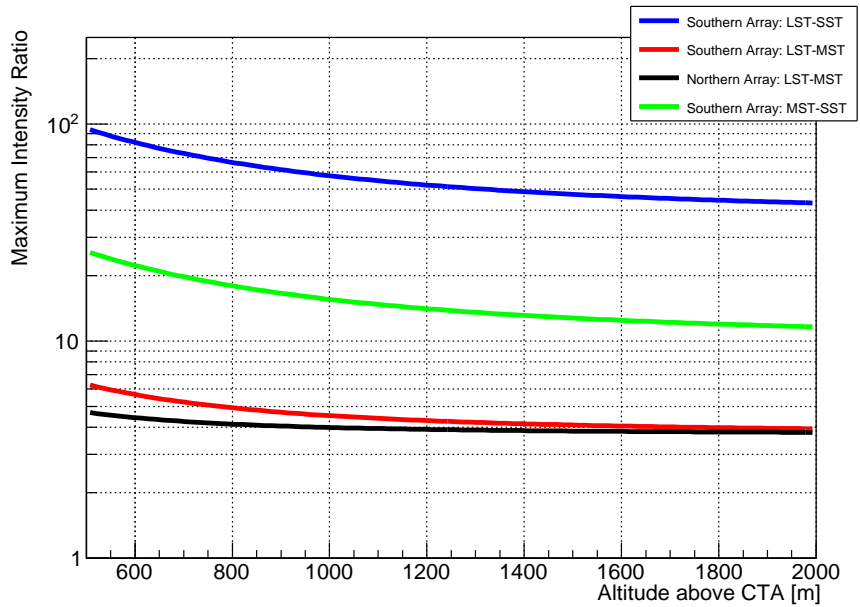}
  \caption{The ratio of the light intensity, as given by Eq. 5, observed by the telescope closest to the UAV compared to the furthest telescope, as a function of UAV altitude above CTA, when comparing between different types of telescopes. The array layout is taken from \cite{layout}. Since we are considering the maximal difference in light intensity across the array, for all cross-calibrations, the larger telescope is assumed to be the closest telescope. Note that, contrary to Figure 4, a log scale has been used here.}
  \label{intendiff}
\end{figure*}

The maximum differences in the recorded signal amplitude, when comparing between the different telescope sizes of SST, MST and LST, are shown in Figure \ref{intendiff} as a function of UAV altitude above the array assuming that the UAV is flown directly over the centre of the array. As before, since we are considering the maximal difference in light intensity across the array, for all cross-calibrations the larger telescope is assumed to be the closest telescope. To account for the additional reflectivity losses of the dual-mirror SST, assumed to be 10\% at worst, the intensity of light seen by the SSTs is calculated to be 90\% of Eq. 5 when compared to the single dish MST and LST designs. 

As before, Figure \ref{intendiff} indicates a preference towards higher UAV operating altitude for minimising the difference in signal amplitude across the array. However, the difference in mirror area for the various telescope types results in an irreducible offset in the maximum difference in recorded signal amplitude. The smallest difference in mirror area is between the LSTs and MSTs with 387~m$^2$ and 100~m$^2$ respectively. In Figure \ref{intendiff}, at the reference altitude of 1300~m above CTA there is a factor of $\sim4$ difference in the light intensity focussed onto the focal plane for LST and MSTs, when considering one UAV positioning for the northern array. For the larger southern array, the LST and MST intensity ratio is slightly larger due to the larger distances between the nearest and furthest telescope. Since there is a factor of $\sim10$ difference in the mirror area between the MSTs and SSTs, the maximum recorded signal ratio across CTA's southern array is $\sim15$. The difference between LSTs and the SSTs is always larger than 45 for the $500-2000$~m altitude range considered here. For the LST and MSTs, the dynamic range of the cameras is expected to be in the range of $1-2000$ photoelectrons, though for some camera prototypes, this dynamic range will be covered by two gain channels. As such, for cross-calibration, comparing the LSTs to the MST telescopes and the MSTs to the SST telescopes can be achieved by selecting the calibration light intensity such that at least $50-100$ photoelectrons are detected by all telescopes, without saturating the camera closest to the UAV. Furthermore, if the calibration payload is capable of a range of illumination intensity levels, a cross-calibration of the LST and SST components of CTA can only be achieved by using the MSTs as a common reference.
  
It is important to note however, that the cross-calibration of the LST and SST components of CTA by using the MST telescopes as a reference, requires a longer UAV flight time since two illumination levels are needed and there will be an increase in the systematic uncertainty of the technique. As such, the possibility of directly comparing the LST and SST signals should to be considered. Across CTA, the largest difference in mirror area is between the LSTs and SSTs, with a factor of $\sim40$ difference in the mirror area between the two designs. This large mirror area difference, when convolved with the geometrical differences associated with the array layout, results in a maximum ratio difference of $\sim53$ in the recorded signal amplitude for one UAV positioning at an altitude of 1300~m. As such, requiring at least $50-100$ photoelectrons in all SST telescopes when comparing the LST and SST telescopes is not possible since the LSTs will be saturated by the calibration signal. While cross-calibrating the LSTs and SSTs is possible if the minimum required number of photoelectrons observed by all telescopes is reduced, the requirement that a Gaussian approximation of the Poissonian photoelectron statistics is still appropriate, requires an average signal of at least 30 photoelectrons for the SSTs resulting in $\sim1800$ photoelectrons for the LSTs. Both of these signals are at the edge of the allowed dynamic range of signals and as such, methods for reducing the maximum intensity difference across CTA should be considered when cross-calibrating the SST and LST components of CTA is necessary.

One such method would be to take advantage of the active mirror control (AMC) system present in all CTA telescopes. In the case of the LSTs, for example, the AMC system allows for real-time changes to the alignment of individual mirrors to correct for the deformations in the LST dish and camera support due to wind and gravitational loads \cite{hayashida}. During a UAV flight, the AMC system can quickly and easily change the focus of the LSTs and slightly de-focus the calibration light spot such that it falls on a group of seven pixels, instead of just one. This would reduce the intensity of the light seen in the LST's central pixel by a factor of seven, with the calibration signal being extracted by summing over the seven central pixels. This approach however will result in an increased contribution to the extracted signal from background light and may be sensitive to cathode nonhomogenity for the PMT-based LST and MST cameras. 

\begin{figure*}
 \begin{minipage}{140mm}
  \centering
\includegraphics[width=.65\textwidth]{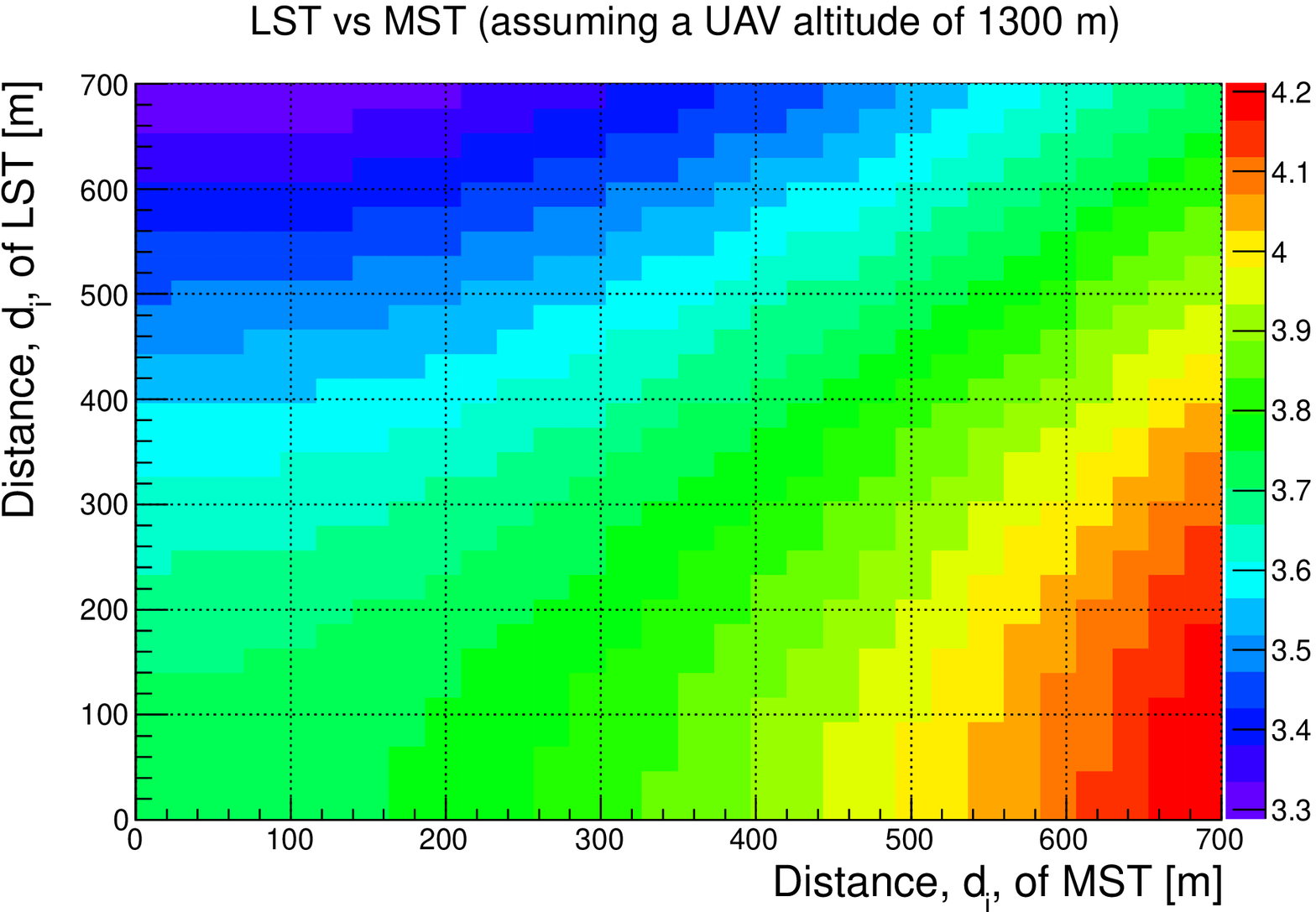}\hfill
\includegraphics[width=.65\textwidth]{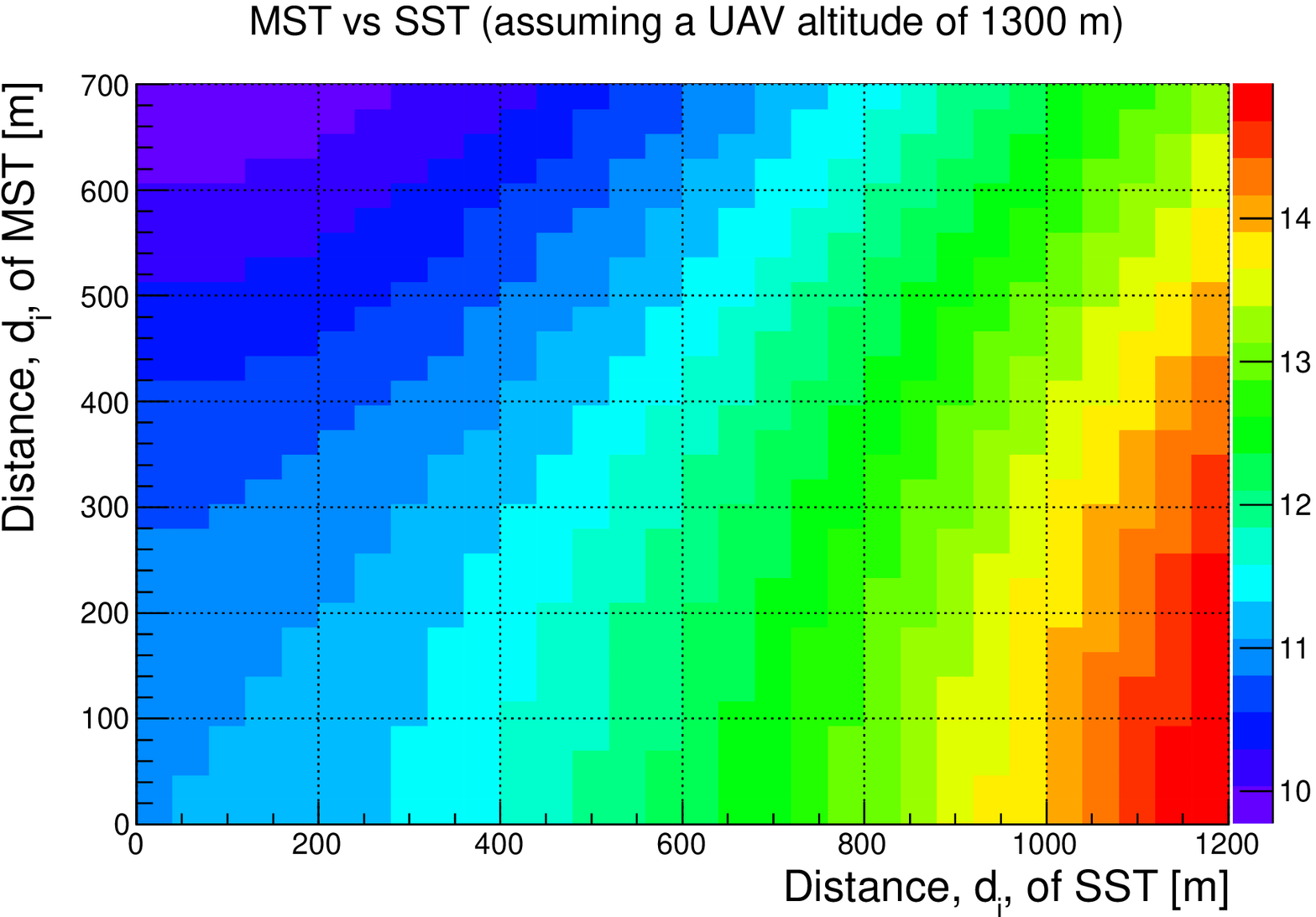}\hfill
\includegraphics[width=.65\textwidth]{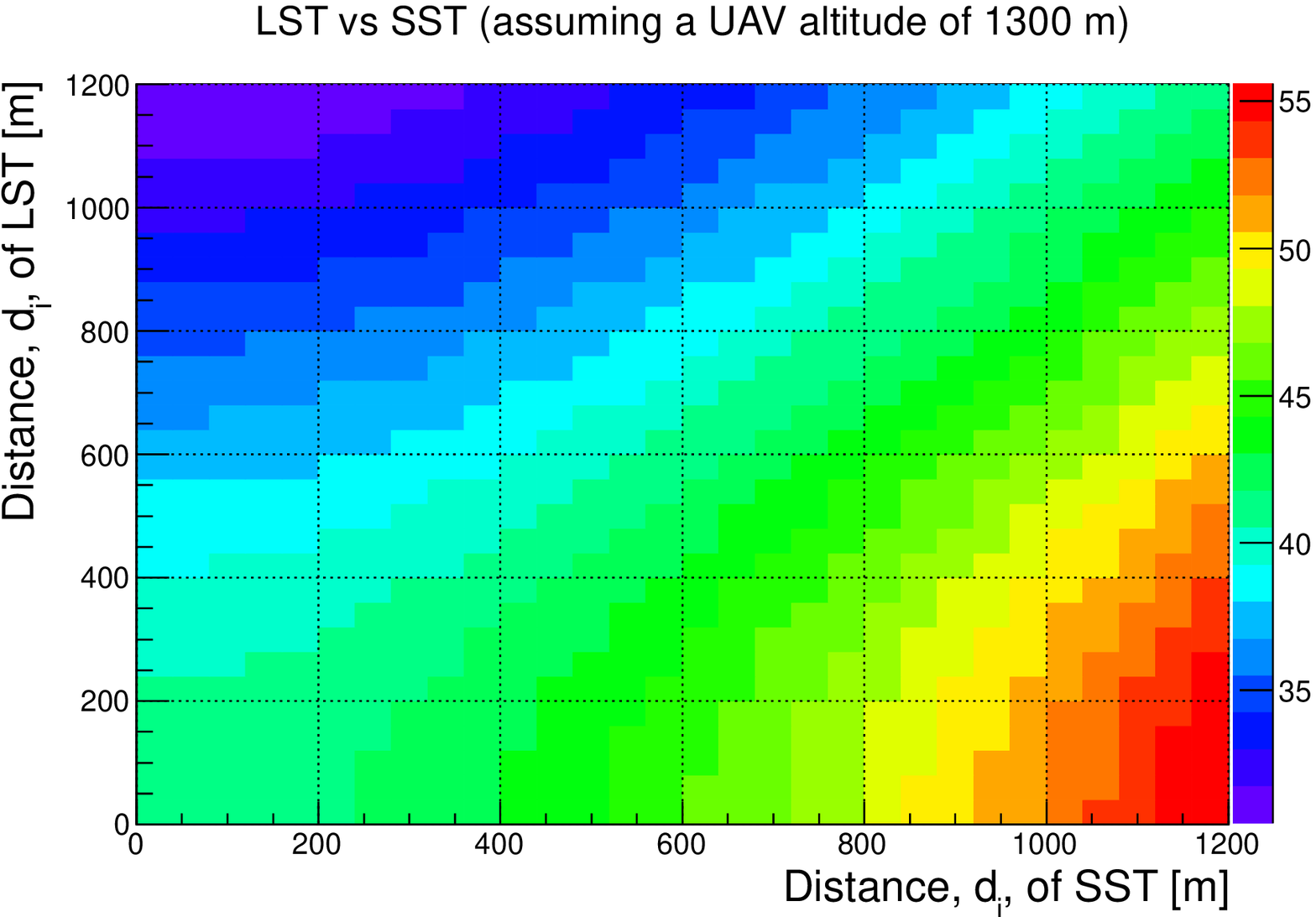}
\caption{Assuming that the UAV is flown at an altitude of 1300~m above CTA, the maximum signal intensity ratio was calculated with the horizontal distances of the individual telecopes from the UAV, $d_i$ in Figure 1, considered to be free parameters. The resultant ratios are indicated by the colour scale for \textit{Top:} LST-MST cross-calibration, \textit{Middle:} MST-SST cross-calibration and \textit{Bottom:} LST-SST cross-calibration.}
 \label{optimise}
 \end{minipage}
\end{figure*}

Another possibility to reduce the maximum intensity across the array when cross-calibrating the SSTs and LSTs is to illuminate CTA with multiple UAV positionings above the array. To optimise these UAV positionings with a view of minimising the maximum intensity ratio, the UAV was assumed to be flown at the reference altitude of 1300~m above the array, while the horizontal distances of the individual telecopes from the UAV, $d_i$ in Figure 1, were considered to be free parameters. The resultant intensity ratios for the LST-MST, MST-SST and LST-SST comparisons can be seen in Figure \ref{optimise}, where we see that the intensity ratio is smallest when the horizontal distance between the smaller telescope in a given cross-calibration and the UAV tends to zero, while the horizontal distance between the larger telescope is as large as possible (which will be defined by the eventual layout of the CTA). As such, to investigate the effect that multiple UAV illumination positions have on the maximum intensity ratio, four UAV positions are considered positioned equally spaced around the edge of CTA, in addition to a single UAV positioning above the centre of the array. The choice of four positionings around the edge of the array represents a `conservative' choice for the multiple positions, the number and locations of the multiple positionings will be reconsidered once the final topological layout of the CTA telescopes of the southern array is known.

\begin{figure*}
 \centering
  \includegraphics[width=120mm]{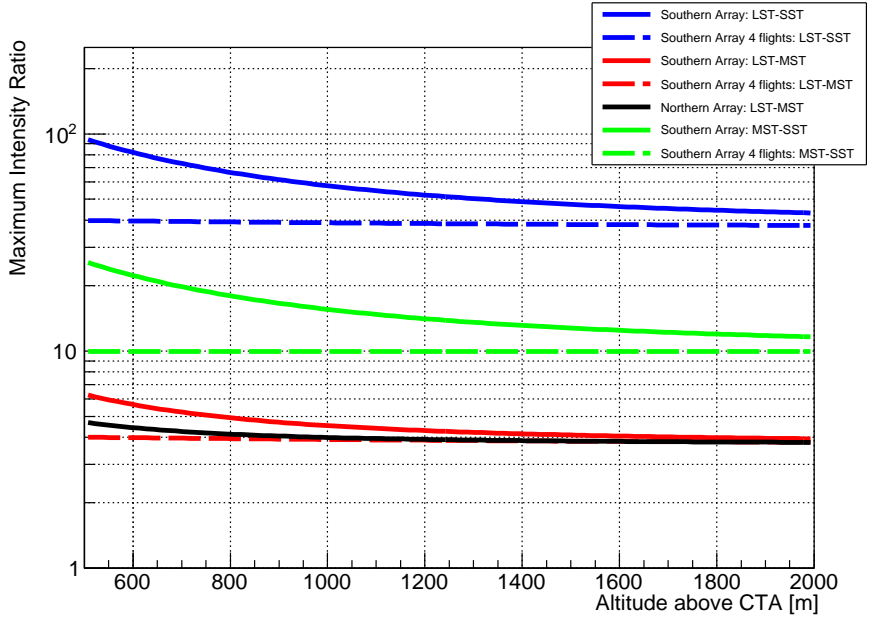}
  \caption{The ratio of the light intensity, as given by Eq. 5, observed by the telescope closest to the UAV compared to the furthest telescope, as a function of UAV altitude above CTA, when comparing between different types of telescopes. The array layout is taken from \cite{layout}. The solid lines represent the ratios when only one flight is conducted directly above the centre of CTA (as in Figure \ref{intendiff}), with the dashed lines representing four flights located at the edge of the array.}
  \label{intendiff2}
\end{figure*}

The resultant maximum light intensity ratios, as a function of altitude above CTA can be seen in Figure \ref{intendiff2} where the solid lines represent the intensity ratios when considering one UAV positioning above the centre of the array, and the dashed lines represent the ratio for the multiple flights. Again, the largest telescope is assumed to be the telescope closest to the UAV. Assuming four UAV positionings around the edge of the array reduces the intensity ratio, with the magnitude of the reduction being dependent upon the telescope size classes that are being cross-calibrated and the altitude of the UAV. Importantly, when flying at the edge of the array, there is no strong dependency of the intensity ratios on the UAV flight altitude. Nonetheless, when comparing to the one flight ratios we see that, for cross-calibrating the MST and SST telescopes on the southern array, the ratio drops to $\sim10$, while the intensity ratio when cross-calibrating the LST and MST telescopes drops to $\sim4$. Importantly, for the LST and SST cross-calibration, at the reference altitude of 1300~m above CTA, the ratio is reduced to $\sim38$. This reduction in the intensity ratio allows us to cross-calibrate all SST telescopes with a minimum signal of $50$ photoelectrons without saturating the LSTs. This minimum signal is sufficiently large to allow a Gaussian approximation of the photoelectron statistics for the SSTs, while making sure that the LST signal does not suffer from any possible non-linear camera response associated with saturation.

\section{Achievable accuracy}

The accuracy with which a UAV-based calibration instrument can cross-calibrate CTA, whether it be an absolute or relative calibration, is governed by a relatively small number of key parameters. These parameters are as follows: 

\begin{itemize}
 \item \textbf{Photoelectron statistics:} For a Gaussian approximation of the recorded photoelectron statistics to be sufficiently precise, each telescope should receive at least 30-50 photoelectrons per calibration flash. As discussed in \textsection \ref{secinter}, a maximum intensity ratio of $\sim38$ is expected when cross-calibrating the LST and SST telescopes and a ratio of $\sim10$ is expected when cross-calibrating the MST and SSTs. As such, for a single calibration flash, the furthest SST records $\sim50$ photoelectrons, while the closest MST and LST records $\sim500$ and $\sim1900$ photoelectrons respectively. This equates to a $\sim14$\%, $\sim4$\% and $\sim2$\% statistical uncertainty for the SSTs, MSTs and LSTs respectively. However, taking 400 calibration flashes, for example, will reduce the mean statistical uncertainty to $<1$\% for all telescope types. 

 \item \textbf{Light source intensity:} The calibration light source used for the cross-calibration is envisaged to be modelled on the Gamma-ray Cherenkov Telescope's (GCT) flasher calibration system \cite{me}. This system consists of an array of UV emitting Light Emitting Diodes (LEDs) behind a circular diffuser to create a wide-beamed calibration pulse with a FWHM pulse duration of $\sim4$ ns \cite{me}. The lightweight, small form-factor and inexpensive nature of LEDs makes them an attractive possibility for a UAV-based calibration light source. While the light output of LEDs is temperature dependent, characterisation of GCT's flasher calibration system has found that a temperature stability of $\pm 2^{\circ}$C for the LED is sufficient for the systematic error on the number of photons emitted per pulse and the pulse FWHM to be $\leq4$\% \cite{me}. This systematic error can be further reduced to the level of 1\% with calibration payload temperature stabilisation and monitoring, and the use of an onboard photodiode (e.g. \cite{minos,bauml}). The thermal protection will be in the form of insulation in the aerodynamic fairing around the calibration payload, heating elements and a thermometer to measure the ambient temperature within the calibration payload fairing.

 \item \textbf{Absolute UAV position:} Uncertainty in the absolute position of the UAV with respect to CTA introduces a systematic error in both the intensity of the light incident on the individual telescopes and the atmospheric optical depth associated with aerosol and molecular extinction. Commercially available Global Navigation Satellite Systems (GNSS), flown in a UAV, have been found to have an uncertainty of $\pm1$~m in regions where a satellite-based augmentation system (SBAS) is available \cite{me2,hayashi,matthews}. With the UAV at a reference altitude of 1300~m, an uncertainty of 1~m in the altitude of the UAV equates to a maximum uncertainty of $\sim0.2$\% in the intensity of the calibration light source at ground level, while a 1~m uncertainty in the horizontal position of the UAV equates to a maximum uncertainty of $\ll 0.1$\% in the intensity of the calibration light source at the telescopes. Furthermore, at an operational altitude of 1300~m, with a horizontal positional uncertainty of 1~m, the UAV will remain in the field of view of a single pixel for all the proposed telescope cameras. With regards to the atmospheric transmission of the calibration signal, a 1~m uncertainty in the altitude or horizontal position of the UAV results in a $\ll 1$\% difference in the atmospheric transmission, $\mathscr{T}_{atm}$, as defined by Eq. 3. It should be noted however, that SBAS is not globally available. While CTA-North will be covered with a SBAS system, CTA-South will not be. To offset the negative impact of no SBAS coverage for CTA-South, the use of a Real-Time-Kinetic (RTK) system, with a positional accuracy $<5$~cm \cite{hayashi}, will be considered. It is worth highlighting that, in addition to minimising the contribution to the global systematic uncertainty from the absolute UAV position, the cm-level accuracy will reduce battery usage, due to a smaller number of position corrections.

 \item \textbf{UAV 3D position stability:} The ability of the UAV to hold a given position with respect to CTA, with as little movement as possible, is also an important consideration for the cross-calibration accuracy. This is primarily governed by the amount of wind loading on the UAV and the mass distribution around the UAV's centre of mass. A UAV payload with a cross-sectional profile area as little as 100~cm$^2$ exposed to a 10~m$/$s wind load will experience a 30 N force parallel to the wind direction. Furthermore, negative pressures caused by airflow vortices around the payload will create forces perpendicular to the wind direction. The magnitude and varying direction of these forces result in the UAV experiencing a large amount of buffeting that consequently increases the positional uncertainties of the UAV's absolute position. Correcting for this buffeting also greatly increases the battery usage, thus reducing the total flight time. As such, to maximise the positional stability of an airborne calibration system, as well as the total flight time, attention must be given to the aerodynamic profile of the scientific payloads.

 \item \textbf{UAV roll, pitch and yaw stability:} The rotational stability of the UAV around the x, y and z axes, referred to as roll, pitch and yaw respectively, is primarily governed by the amount of wind loading on the UAV. Assuming a circular light pool from the calibration light source, a yaw instability does not introduce an additional systematic uncertainty to the cross-calibration of CTA. Assuming the use of a circular diffuse with a uniform intensity across the considered solid angle \cite{me}, an uncertainty in the roll or pitch of the UAV will simply translate to an uncertainty in the position of the calibration light pool on the ground. The effects of the roll and pitch uncertainty can be mitigated however if a reduced illumination area is considered for calibration. 

 \item \textbf{Atmospheric transmission:} As discussed in \textsection \ref{crosscalib}, atmospheric transmission is a summation of both molecular and aerosol transmission, with the aerosol transmission dominating over the molecular component for astronomically clear nights. Molecular extinction can be calculated analytically if the vertical pressure, temperature and humidity profiles of the atmosphere are known. These profiles are usually recorded through intense radiosonde campaigns measured over many years (e.g. \cite{augerradiosonde}). However, housing temperature, humidity and pressure sensors with the calibration light source onboard the UAV allow for in-situ profiling of the atmosphere\footnote{It is worth highlighting that housing temperature, humidity and pressure sensors onboard the UAV will not only allow us to minimise the systematic uncertainty associated with atmospheric transmission, it will also allow us to conduct a radiosonde measurement of the first kilometre of the atmosphere above CTA with each calibration flight.}. These real-time measurements allow us to calculate the molecular extinction to an accuracy of $<1$\% and hence the transmission term $\mathscr{T}_{mol}$ to $\ll 1$\%, even if UV light is used. The magnitude of the aerosol extinction depends upon the density and shape of the scattering particle. The Pierre Auger Observatory found the vertical aerosol optical depth at an altitude of 1 km above their detector to be $\sim$0.01 on astronomical clear nights \cite{augerclf2}, though this increases by an order of magnitude for nights with high dust content. If the UAV is operated in conjunction with an instrument able to quantify the atmospheric aerosol content, such as a Raman LIDAR (eg see \cite{gaug2}) or a nephelometer onboard the UAV\footnote{The dust content in the first kilometre of atmosphere above the ground is poorly currently constrained. With the advent of low mass, hand-held nephelometers, housing a nephelometer on the UAV to directly sample the dust content during each flight is now a possibility.}, an uncertainty of $<1$\% for the aerosol transmission from 1.3~km to the ground can be assumed.   

 \item \textbf{Background light:} As discussed in \textsection\ref{secintra}, the expected background flux, $F_{bgr}$, for the LST, MST and SST size classes is on the order of 2.0, 0.9 and 0.15 photoelectrons respectively per pixel for a 5 nanosecond calibration pulse. A maximum intensity ratio of $\sim38$ is expected when cross-calibrating the LST and SST telescopes, with the furthest SST structure recording $\sim50$ photoelectrons without saturating the closest LSTs. As such, assuming direct cross-calibration of the LSTs with the SSTs, the background light amounts to $\sim0.6$\% of the UAV signal for the SST. This drops to $\sim0.3$\% for the SST when cross-correlating the MST and SSTs (due to the larger allowed signal of $\sim100$ photoelectrons). For the MST, the background light amounts to $\sim2$\% of the UAV signal when cross-calibrated with the LSTs, which drops to $<0.1$\% when cross-calibrating against the SSTs. For the LSTs, the background light always amounts to $<0.3$\% of the UAV signal. 

 \item \textbf{Telescope depth of field:} During calibration runs, the UAV will be located in an altitude range of $\mathcal{O}(1-1.5)$~km above CTA, while the EAS that CTA will observe are $\mathcal{O}(8-10)$~km above the array. The proximity of the UAV relative to the distance at which the CTA's telescopes are focussed may result in a smearing of the UAV-based calibration light source due to aberration effects. The magnitude of the abberation effects is dependent upon the depth of field for the individual telescopes and the position of the UAV relative to the telescope's optical axis. At its worst, the point-like image of the calibration light source will be smeared across several pixels, thus increasing the contribution from background light to the recorded signal. Given the pixel FoVs for the different CTA telescope designs, ranging from $0.25^{\circ}$ for the Davies-Cotton SST design \cite{sst}, to $0.067^{\circ}$ for the SCT \cite{sct}, detailed ray-tracing simulations are required for all CTA telescope designs to quantify the systematic error associated with aberration effects. Using the \textsc{robast} ray tracing software \cite{robast}, simulations for the GCT, showed $\sim0.1^{\circ}$ difference in the point spread function (PSF) across the camera field of view \cite{gctray}. Critically the PSF was found to reduce for objects located at distances less than 10~km from the telescope, when they are viewed at small field angles. On the other hand, the abberation effects associated with the parabolic mirror of the LST, when focussed to observe EAS, will smear the calibration signal from a UAV located 1~km from the telescope across more than ten pixels. Likewise for the MST prototype, the calibration signal from a UAV at an altitude of 1~km above the array will be smeared across five pixels. However, for telescope with automated mirror alignment systems, any smearing of the calibration signal can quickly be corrected for by refocussing the mirrors.

 \item \textbf{Camera flat-fielding:} To minimise the systematic uncertainty of the cross-calibration, the same pixel should be illuminated for each calibration flight. If this is not achieved, an additional systematic uncertainty is introduced by converting the observed signal, as measured by the illuminated pixels, to a calibration signal as calculated for a camera-averaged pixel sensitivity which is derived from the camera flat-fielding system\footnote{It is worth noting that the flexibility of the UAV affords us the possibility of also illuminating different parts of the camera, thus exploring different regions of the telescope's optical system.}. The magnitude of this additional uncertainty depends upon how different the sensitivity of the illuminated pixel is compared to camera-averaged pixel sensitivity and is primarily governed by the difference in incidence angles for light from a telescope mounted flat-fielding system compared to the angles expected for a UAV based calibration system and temporal variability in the camera's response. Undertaking UAV cross-calibration simultaneously with the on-telescope flat-fielding calibration procedure will remove the latter effect, while detailed studies of the angular dependence of camera sensitivity is needed to quantify the contribution from the former. While detailed simulations are needed to determine how a single pixel's signal translates to a camera-averaged signal, the magnitude of this uncertainty is on the order of $5-7$\%.


 \item \textbf{Telescope point spread function:} The optical systems of each telescope size class have different PSF. Combine this with the pixel FoV for the individual camera designs envisaged for CTA results in each telescope design focusing different percentages of the calibration light onto the illuminated pixel. Unless they are corrected for, these differences can result in an additional systematic uncertainty when cross-calibrating different telescope size classes. Accounting for this additional uncertainty is limited by the accuracy with which we can determine the amount of light lost out of the illuminated pixel and as such, requires detailed ray tracing simulations for each CTA telescope design. Some such simulations have shown the 90\% containment radius to be within the angular size of an individual pixel for all CTA telescope prototypes assuming the light source is viewed along the optical axis of the telescope structure, with some prototype designs having the $\gg99$\% containment radius to be within one pixel \cite{gctray}. Assuming these numbers, the uncertainty associated with correcting for light focussed outside of the central pixel when cross-calibrating between the SST, MST and LSTs size classes is $<5$\% assuming all telescopes are pointing towards the UAV and that there are low wind conditions. This uncertainty is determined from ray tracing simulations of CTA telescope prototypes (e.g. \cite{gctray}). It is important to highlight however, that due to aging of the mirrors, the telescopes' PSFs will degrade over time. As such, ray tracing simulations are needed to ascertain whether this $5$\% uncertainty holds for a variety of mirror reflectivities, over the lifetime of the telescope.

 \item \textbf{Stratification of aerosol distribution}: The atmospheric transmission described in Eq. 4 assumes an atmospheric aerosol distribution that is uniformly stratified above CTA. The stratification of aerosols in the lower atmosphere can occur when the wind speed is low and the wind direction remains temporally stable. These environmental conditions may not be applicable for both CTA-North and CTA-South sites. The assumption of aerosol stratification introduces a maximum systematic uncertainty of 2\% \cite{nolan}.        
\end{itemize}

\begin{table}
\centering
   \caption{Sources of systematic uncertainty and the magnitude of the effect they have on the global systematic uncertainty when using a UAV to cross-calibrate CTA at an altitude of 1300m above the array. For individual sources of uncertainties where there is a range of values, the global uncertainty was calculated assuming the upper bound of the range.}
     \begin{tabular}{lll} \hline 
      Source of Uncertainty	& Inter-calibration     & Cross-calibration  \\ \hline 
      Statistical               & 1\%                   & 1\%                \\ 
      Light source stability    & 1\%                   & 1\%                \\
      Absolute UAV position     & $<$1\%                & $<$1\%             \\ 
      Atmospheric extinction    & 1\%                   & 1\%                \\
      Background light          & $<$0.5\%              & $<$1\%             \\  
      Flat-fielding             & $5-7$\%               & $5-7$\%            \\
      Point Spread Function     & 2\%                   & $<$5\%             \\ 
      Aerosol Distribution      & 2\%                   & 2\%                \\ \hline
      Total Uncertainty         & $6-8$\%               & $8-10$\%           \\ \hline
    \end{tabular}
  \label{uncertainties}
\end{table}

The individual sources of uncertainty for both inter-calibration and cross-calibration are summarised in Table \ref{uncertainties}. For both inter-calibration and cross-calibration, the largest single contribution to the overall systematic uncertainty is converting the optical throughput from one individual pixel to a `camera-averaged' pixel using information from camera flat-fielding. This uncertainty can be minimised if flat-fielding calibration can occur simultaneously with the UAV-based cross-calibration, as is envisaged for CTA observations, and the same pixel is illuminated for each cross-calibration run. The other main contribution to the overall systematic uncertainty is the PSF of the telescopes and the performance of the LED light source. The latter uncertainty can be minimised if the calibration flasher system is characterised in the lab, with appropriate thermal protection to keep the flasher system thermally stable during the flight. Inflight monitoring of the absolute light output from the flasher system with an onboard photodiode will allow this uncertainty to be reduced further.

Another important contribution to the overall error budget is the stability and absolute accuracy of the UAV's position since it effects the amount of atmospheric extinction for a given atmospheric transmission, the flux of the calibration light source incident on the telescopes and whether the calibration light source moves between pixels during the calibration run. While flying at altitudes $>1$ km greatly reduces the effect of the latter two, it increases the uncertainty associated with atmospheric transmission. This increase however is much less than the reduction achieved in the uncertainy associated with the flux received by the telescopes.

For inter-calibration, assuming UAV operation during a camera flat-fielding, with the uncertainty in the UAV's position dominated by the GNSS uncertainty and avoiding pointing the telescopes towards low Galactic and ecliptic latitudes to minimise the amount of background light, an absolute calibration uncertainty of $6-8$\%, including statistical uncertainty, has been determined. For cross-calibrating telescopes from different size classes, with different photosensors, assuming the same precautions are taken for the calibration light source as in done for inter-calibrating telescopes of the same type, the absolute calibration uncertainty increases to $8-10$\%.

\section{CTA Hardware consideration}

Cross-calibration of CTA can be achieved with a UAV-based calibration flasher without saturating the telescope nearest to the UAV. Furthermore, utilising a flashing calibration light source with timing characteristics similiar to the Cherenkov images the CTA telescopes are optimised to observe does not enforce additional hardware requirements for signal amplification and digitisation. Nonetheless, there are still several hardware considerations required of the CTA telescopes for airborne calibration to succeed. These considerations are as follows:

\begin{itemize}
 \item \textbf{Telescope Pointing}: Each telescope should be able to point towards the location of the UAV. For a UAV altitude range of $\mathcal{O}(1-1.5)$~km above CTA, the range of zenith angles at which the UAV will be observed by the telescopes will be $0^{\circ}-27^{\circ}$ for the smaller northern site and $0^{\circ}-50^{\circ}$ for the larger southern site assuming four UAV positionings. As such, all CTA telescope designs will be able to point towards a UAV-based calibration system. 
 
 \item \textbf{Telescope Depth of field}: Ideally each telescope should have a depth of field that allows for a UAV to be located $\mathcal{O}(1-1.5)$~km above the array and for the calibration flashes to be contained within one pixel, as this  allows for a more precise subtraction of the background. If the telescope cannot satisfy this requirement, as is expected for the parabolic mirror of the LST, the use of mirror alignment systems should be considered.

 \item \textbf{Camera Trigger}: Cherenkov telescope cameras usually trigger once a group of neighbouring pixels reaches a pre-defined photoelectron threshold in a given time window to minimise triggering on night sky background (NSB) fluctuations (e.g. \cite{defranco}). Assuming that the depth of field of each telescope type is sufficient for the UAV-based light source to be contained in one camera pixel, each camera must be able to trigger on a single bright pixel.
\end{itemize}

The ability of the camera to trigger on one bright pixel is the primary telescope hardware criterion that needs to be addressed for UAV-based cross-calibration. The minimum expected average signal of 50 photoelectrons, as defined in \textsection\ref{secinter}, compared with the maximum expected background of a few photoelectrons, as defined in \textsection3, suggests the possibility of false triggers from NSB fluctuations is small, though the possibility of stars within the field of view triggering the camera needs to be investigated once individual camera designs have been finalised. To facilitate camera triggering, a dedicated programmable trigger might be considered. The simpliest such trigger would be to blind the entire camera and only trigger on the pixels where the calibration signal is expected. To further reduce the possibility of triggering the camera on NSB fluctuation, this dedicated trigger would also have a higher signal threshold compared to camera operation during science observations. 


\section{Discussion}


The considerable effort that has gone into investigating cross-calibration methods for the CTA telescopes has highlighted the importance of understanding the wavelength-dependence of telescope aging, especially with regards to minimising the systematic uncertainty of the telescopes' optical throughout over the lifetime of CTA. For example, while comparing the images of local muon rings or cosmic ray-induced EAS from individual telescopes allow us to understand the telescopes' optical throughput to $2-3$\% systematic uncertainty, the lack of spectral information for these techniques restricts the accuracy of the method once wavelength-dependent degradation of the telescope's optical systems occurs \cite{gaug2}. 

Accurately accounting for possible wavelength-dependence of the optical systems' degradation requires a cross-calibration device or methodology with MWL capability. To address the uncertainty associated with wavelength-dependent degradation, a MWL-capable vertically-firing CLF has been investigated \cite{markusclf}. However, while a CLF would be able to achieve a relative cross-calibration of CTA to a 5\% level of uncertainty, the hardware requirements to use a CLF, particular with regards to signal digitisation and camera readout, appear to be prohibitive due to the unacceptably long calibration signal pulse widths. A flashing calibration system that has temporal characteristics similiar to the Cherenkov radiation produced by EAS does not impose such strict hardware requirements. Adapting current LED-based flasher calibration systems, which operate at UV wavelengths \cite{me,hanna}, to operate at other wavelengths allows for a MWL flasher system with temporal characteristics similiar to the Cherenkov radiation emitted by EAS. Housing such a MWL-capable flasher system on a UAV allows us to perform a MWL cross-and inter-calibration of CTA, and still achieve a total systematic uncertainty in the range of $6-8$\% and $8-10$\% respectively, even after wavelength-dependent degradation of the telescopes' optical system has occurred. Furthermore, MWL cross-calibration can also be used to correctly account for the effect of MWL degradation on the cross-calibration of CTA with EAS and muon rings, thus providing CTA with several independent and complementary methods to cross-calibrate individual telescopes.


To minimise the contribution to the overall systematic error from background light, cross-calibration with a UAV should be completed during dark observing time. Additionally, to minimise the systematic uncertainties of the cross-calibration, all telescopes are required to point to the UAV. As such, an important consideration for a UAV-based system is the time-taken to complete the calibration since it defines how much potential observing time is spent calibrating the array. The operational trigger rate for calibration systems currently considered for CTA, such as that for the GCT \cite{me}, suggest that the time needed to illuminate CTA with sufficient photon statistics, at a given UAV position, is of the order of a minute per wavelength, assuming a 10 Hz calibration flash frequency. As such, the dominant contribution to the time taken to complete the airborne calibration is the movement of the UAV from launch to the illumination positions and the subsequent landing. For the northern array, all telescopes can be cross-calibrated with one positioning above the centre of the array, assuming an operational altitude of 1300~m above the array. The rate of climb performance for current off-the-shelf UAV platforms implies that the cross-calibration of the northern array will take as little as eight minutes. For the southern array, all telescopes will be illuminated with five UAV positionings at an altitude of 1300~m above the array. The flight performance, such as rate of climb and maximum cruising speed, of current off-the-shelf UAV platforms means that the complete cross-calibration will take as little as $\sim30$ minutes, assuming one take-off$/$landing cycle. A 30 minute flight time is possible with current LiPo battery technology (assuming a $30$ Ah battery capacity). If the cross-calibration of the southern array is undertaken by five UAVs, with a single UAV considered for each illumination position, this 30 minute period can be reduced to eight minutes when all five UAVs are simultaneously climb to their predefined GNSS positions above the array. However, if the scheduling of the UAV calibration flight is optimised, such that the UAVs take-off during twilight and are in position above the array at the start of the dark time, the time needed for the UAV-based calibration can be reduced further. Once dark time starts, and the PMT HVs have stabilised, the UAV-based cross-calibration can commence. In this approach the time needed for the UAV to complete the cross-calibration can effectively be minimised to the amount of time needed to flash the telescopes. Once the calibration flashes are completed, and the UAV starts it's decent, the telescopes can slew to the first science target of the night.

The small temporal period required by a UAV-based system to cross-calibrate CTA is also beneficial with regards to operating costs. While a UAV-based system is expected to have two operators during each calibration flight, one operator as a pilot and the other as a payload specialist, the small amount of time taken to complete the cross-calibration of the entire array limits the financial impact of needing two people to complete the calibration.

Finally it is worth noting that, in addition to the primary payload of a MWL-capable flasher system for cross-calibration, a secondary environmental payload of nephelometer, temperature, humidity and pressure sensors allows us to minimise the systematic uncertainty associated with atmospheric transmission. Information from such a characterisation allows us to more accurately account for atmospheric extinction of Cherenkov photons, thus improving the systematic uncertainty on the energy scale of CTA \cite{nolan}. Furthermore, the dust content in the first few hundreds of metres of atmosphere above ground-level is currently poorly constrained by classical LIDAR systems. Taking advantage of the flexibility of a UAV-based calibration system by housing a nephelometer onboard the UAV to perform dedicated dust-sampling flights allows us to map the distribution of dust above CTA. This information can be included in the simulations of propagation of Cherenkov photons through the lower atmosphere, thus improving the energy resolution of CTA. 

\section{Conclusions}

This feasibility study has found that the cross-calibration of CTA is possible with a UAV-based airborne calibration system. With a number of assumptions on the atmospheric conditions and performance of the calibration light source, this cross-calibration was found to achieve a $6-8$\% uncertainty when calibrating telescopes of the same size class, and an uncertainty of $8-10$\% when cross-calibrating telescopes of different size class. The difference between the two uncertainties is primarily driven by the different telescope's PSF when comparing telescopes of different size classes. Irrespective of whether it is inter- or cross-calibration, the overall largest contributor to the error budget for UAV-based calibration of optical throughput is the camera flat-fielding, that is, how the UAV-illuminated pixel relates to the average camera pixel sensitivity. 

Equipping the UAV with a MWL-capable flasher system affords the possibility to monitor the wavelength-dependent degradation of CTA telescopes' optical system. Not only does this allow us to maintain this $6-10$\% uncertainty after the first few years of telescope deployment, it also allows us to accurately account for the effect of MWL degradation on the cross-calibration of CTA with EAS and muon rings, thus providing CTA with several independent and complementary methods of cross-calibrating individual telescopes.

In addition to the primary light source calibration payload, housing environmental sensors on the UAV system allows us to not only minimise the systematic uncertainty associated with atmospheric transmission of the calibration signal, it also allows us to map the dust content above CTA as well as monitor the temperature, humidity and pressure profiles of the atmosphere above CTA. This indepth knowledge of the atmosphere above CTA will allow us to improve the energy resolution of CTA in addition to minimise the systematic uncertainty of the UAV-based cross-calibration technique.

Finally, it is important to highlight that a MWL-capable UAV platform is currently the only device on the CTA calibration road-map capable of simultaneously monitoring the MWL performance of numerous CTA telescopes quickly.

\vspace{5mm}

\textbf{Acknowledgments}
I thank Anthony P. Brown for his helpful discussions and insights, particularly with regards to flight dynamics, and Nicholas J. Brown for his comments on wind loading, both of which have been invaluable to this paper. I would also like to thank Markus Gaug, Felix Werner, Paula Chadwick, Michael Daniel, Cettina Maccarone and Dan Parsons for fruitful conversations. I also thank the referee for their comments that improved the quality and clarity of this paper. This work was undertaken with the financial support of Durham University and a UK Science and Technology Facilities Council consolidated grant ST/L00075X/1. This paper has gone through internal review by the CTA Consortium.

\end{document}